\documentclass[twocolumn,final]{elsarticle}

\usepackage{amsmath}
\usepackage{amssymb}

\usepackage{xcolor}
\usepackage{hyperref}
\hypersetup{linktocpage, colorlinks=true, linkcolor=blue, citecolor=blue}
\usepackage{graphicx}
\usepackage{makecell}
\usepackage{subcaption}
\usepackage{balance}
\usepackage{float}

\begin{document}

\begin{frontmatter}

\title{Hyperbolic Graph Generator}

\author[add1]{Rodrigo Aldecoa\corref{cor1}}
\ead{raldecoa@neu.edu}
\author[add2]{Chiara Orsini}
\ead{chiara@caida.org}
\author[add1,add3]{Dmitri Krioukov}
\ead{dima@neu.edu}

\cortext[cor1]{To whom correspondence should be addressed}
\address[add1]{Northeastern University, Department of Physics, Boston, MA, USA}
\address[add2]{Center for Applied Internet Data Analysis, University of California San Diego (CAIDA/UCSD), San Diego, CA, USA}
\address[add3]{Northeastern University, Department of Mathematics, Department of Electrical\&Computer Engineering, Boston, MA, USA}

\begin{abstract}
  Networks representing many complex systems in nature and society share some common structural
  properties like heterogeneous degree distributions and strong clustering. Recent research on
  network geometry has shown that those real networks can be adequately modeled as random geometric
  graphs in hyperbolic spaces. In this paper, we present a computer program to generate such graphs.
  Besides real-world-like networks, the program can generate random graphs from other well-known
  graph ensembles, such as the soft configuration model, random geometric graphs on a circle, or
  Erd\H{o}s-R\'enyi random graphs. The simulations show a good match between the expected values of
  different network structural properties and the corresponding empirical values measured in
  generated graphs, confirming the accurate behavior of the program.
\end{abstract}

\end{frontmatter}

\noindent
\textbf{Program summary}

\begin{small}
\noindent
{\em Program title: Hyperbolic Graph Generator}               \\
{\em Program summary URL: \url{http://named-data.github.io/Hyperbolic-Graph-Generator/}}\\
{\em Licensing provisions: General Public License, version 3}  \\
{\em Programming language: C++}                                   \\
{\em Computer/Operating system: Any}                                   \\
{\em Nature of problem: Generation of graphs in hyperbolic spaces.}      \\
{\em Solution method: Implementation based on analytical equations.}     \\
{\em Running time: Depends on the number of nodes. A few seconds for graph in the example provided.}  \\
{\em Other features: Can be used as a command-line tool or installed as a library to support more complex software.}
\end{small}

\section{Introduction}

The interactions between components of a complex system are often represented as a network. This
modeling allows for rigorous mathematical treatment, and broadens our understanding of the
system~\cite{Dorogovtsev2010}. Many real networks possess common structural
patterns, including, in the first place, heterogeneous (often power-law) distributions
of node degrees~\cite{Barabasi1999}, and strong clustering, i.e., higher numbers of triangular
subgraphs than predicted by classical random graph models~\cite{Newman2010}.
Recently introduced geometric graph models, based on the assumption that nodes in real networks are
embedded in latent hyperbolic spaces~\cite{Krioukov2010,Papadopoulos2012},
reproduce these common structural properties of real networks. Furthermore, these hyperbolic graphs
replicate dynamical processes on top of real networks~\cite{Boguna2010} and accurately predict missing
links in them~\cite{Papadopoulos2014}.

\begin{table*}[ht]
\centering
\begin{tabular}{|c|c|c|c|}\hline
\diaghead{zzzzzz}%
{$\gamma$}{$T$}&
$0$&$(0,\infty)$&$\infty$\\\hline
$[2,\infty)$&Hyperbolic RGGs&Soft hyperbolic RGGs&Soft configuration model\\\hline
$\infty$&Spherical RGGs&Soft spherical RGGs&Erd\H{o}s-R\'enyi\\\hline
\end{tabular}
\caption{
Regimes in the model.
\label{tab:status}
}
\end{table*}

In this work we present a program to generate random hyperbolic graphs. This software implements and
extends the network model introduced in~\cite{Krioukov2010}. Nodes are randomly sprinkled on a
hyperbolic disk, and the probability of the existence of an edge (the connection probability) between
two nodes is a function of the distance between the nodes in the hyperbolic space. Thus generated
graphs have strong clustering, and node degree distributions in them are power laws.
Moreover, other popular and well-studied random graph ensembles, namely the soft configuration
model (SCM)~\cite{Park2004}, (soft) random geometric graphs (RGGs) on a
circle~\cite{Penrose2003,Penrose2013}, and Erd\H{o}s-R\'enyi (ER) random graphs~\cite{Erdos1959},
appear as degenerate regimes in the model. Table \ref{tab:status} shows all the model regimes,
the total of six. Each regime is defined by the values of only two parameters: $\gamma$, which is
the expected exponent of the power-law degree distribution, and temperature $T$,
the parameter controlling the strength of clustering in the network.

Researchers in different disciplines may benefit from the use of random hyperbolic graphs in their work.
Yet the full implementation of the model and all its regimes is a tricky business, which involves dealing with some delicate details,
due to a variety of internal parameters and their interactions over the six regimes. In Sections~\ref{sec:implementation}-\ref{sec:infinite_gamma} we describe
the implementation details of the model, including how all the parameters are calculated in each regime. A good match between the values of expected
graph properties and their observed values in generated graphs is confirmed in Section~\ref{sec:simulations}.

\section{Input parameters and coordinates}\label{sec:implementation}
\subsection{Input parameters}
The program input parameters are the number of nodes $N$, the target expected average degree $\bar{k}$
of the network, the target expected power-law exponent $\gamma$ of the degree distribution, and
temperature $T$. The combination of $\gamma$ and $T$ values will define the graph ensemble from
which generated networks are sampled (Table~\ref{tab:status}).

Given the input parameters, the graph generation process consists of three steps:
\begin{enumerate}
\item Compute the internal parameters, such as the radius $R$ of the hyperbolic disk occupied by
  nodes, as functions of the input parameter values, Sections~\ref{sec:finite_gamma},
  \ref{sec:infinite_gamma}.
\item Assign to all nodes their angular and radial coordinates on the hyperbolic plane,
  Section~\ref{sec:coordinates}.
\item Connect each node pair by an edge with probability (the connection probability), which is a
  function of the coordinates of the two nodes, Sections~\ref{sec:finite_gamma},
  \ref{sec:infinite_gamma}.
\end{enumerate}

\subsection{Coordinate sampling}\label{sec:coordinates}
The assignment of node coordinates is done as follows in all the six regimes.

Angular coordinates $\theta$ of nodes are assigned by sampling them uniformly at random from interval $[0, 2\pi)$, i.e., the angular node density is uniform $\rho(\theta) = 1 / (2\pi)$.

Radial coordinates $r \in [0, R]$, where $R$ is the radius of the hyperbolic disk, are sampled from the following distribution, which is nearly exponential with exponent $\alpha > 0$,
\begin{equation}
\label{eq:r_generator}
\rho(r) = \alpha \frac{\sinh \alpha r}{\cosh \alpha R - 1}
\approx \alpha e^{\alpha (r-R)}.
\end{equation}
The calculation of internal parameter $R$ is described in detail below; it is different in different regimes. Internal parameter $\alpha$ depends on the expected exponent $\gamma$ of the power distribution $P(k)\sim k^{-\gamma}$ of nodes degrees~$k$ in generated graphs, and on the curvature of the hyperbolic space $\zeta = \sqrt{-K}$, which is set to $\zeta = 1$ by default. For temperatures $T \leq 1$, this relationship is given by
\begin{equation}
\gamma = 2 \frac{\alpha}{\zeta} + 1,
\end{equation}
while for $T>1$ it becomes
\begin{equation}\label{eq:gamma-hot}
\gamma = 2 \frac{\alpha}{\zeta} T + 1.
\end{equation}
To sample radial coordinates $r$ according to the distribution in Eq.~(\ref{eq:r_generator}), the inverse transform sampling is used: first a random value $U_i$ is sampled from the uniform distribution on $[0,1]$, and then the radial coordinate of node $i$ is set to
\begin{equation}
	r_i = \frac{1}{\alpha}\, \text{acosh}\left( 1 + (\cosh \alpha R -1)\; U_i \right),\quad\text{for $i=1,..,N$.}
\end{equation}

\section{Regimes with finite $\gamma \geq 2$\label{sec:finite_gamma}}

\subsection{$T \in (0,\infty)$: Soft hyperbolic random geometric graphs \label{sec:general_case}}

This is the most general regime in the model, from which all other regimes can be obtained as limit cases. The connection probability in this case is
\begin{equation}
\label{eq:default_px}
p(x) = \frac{1}{1 + e^{\beta (\zeta/2)(x-R)}},
\end{equation}
where $\beta = 1/T$, and $R$ is the radius of the hyperbolic disk occupied by nodes.
The hyperbolic distance $x$ between two nodes at polar coordinates $(r,\theta)$ and $(r',\theta')$ is given by
\begin{align}
x =& \frac{1}{\zeta} \mathrm{arccosh}\left( \cosh \zeta r \cosh \zeta r'\right. \nonumber \\ &- \left. \sinh \zeta r \sinh \zeta r' \cos \Delta \theta \right),
\label{eq:exact_distance}
\end{align}
where $\Delta \theta\ = \pi - | \pi - | \theta - \theta'||$ is the angular distance between the nodes.
To calculate the expected degree of a node at radial coordinate $r$, without loss of generality its angular coordinate can be set to zero, $\theta=0$, so that its expected degree can be written as
\begin{equation}
\bar{k}(r) = \frac{N}{\pi} \int_0^R \rho(r') \int_0^\pi p(x) \, d\theta' dr'.
\end{equation}
The expected average degree in the network is then
\begin{equation}\label{eq:general_case}
  \begin{split}
    \bar{k} &=\int_0^R \rho(r) \, \bar{k}(r) dr \\
    &= \frac{N}{\pi} \int_0^R \rho(r) \int_0^R \rho(r') \int_0^\pi p(x) \, d\theta' dr' dr.
  \end{split}
\end{equation}
Given user-specified values of input parameters $N$, $\beta=1/T$, $\zeta$ and $\bar{k}$, the last equation is solved for $R$ using the bisection method in combination with numeric evaluation of the integrals in the equation. The MISER Monte Carlo algorithm from the GSL library is used to compute the multidimensional integral in Eq.~(\ref{eq:general_case}). The iterative bisection procedure to find~$R$ stops when the difference between the value of the computed integral in Eq.~(\ref{eq:general_case}) and the target value of $\bar{k}$ is smaller than a threshold that is set to $10^{-2}$ by default.

\subsection{Limit $T \to 0$: Hyperbolic random geometric graphs \label{sec:hyperbolic_rggs}}
In the $T \to 0$ ($\beta\to\infty$) limit, the connection probability in Eq.~(\ref{eq:default_px}) becomes
\begin{equation}
\label{eq:Heaviside_f}
p(x) = \Theta (R-x),
\end{equation}
where $\Theta(x)$ is the Heaviside step function, meaning that two nodes are connected if the hyperbolic distance $x$ between them is less than $R$, or they are not connected otherwise.
The expected average degree in the network is given by the same Eq.~(\ref{eq:general_case}), but with $p(x)$ in the last equation.
The value of $R$ is determined using the same procedure as in Section~\ref{sec:general_case}. The only difference is that function $p(x)$ is given by Eq.~(\ref{eq:Heaviside_f}).

\subsection{Limit $T \to \infty$: Soft configuration model}\label{sec:scm}

According to Eq.~(\ref{eq:gamma-hot}), in the $T \to \infty$ limit with finite $\alpha$, to have finite $\gamma$, curvature should also go to infinity, $\zeta \to \infty$,
such that $\eta = \zeta / T$ is finite, and instead of Eq.~(\ref{eq:gamma-hot}) one gets
\begin{equation}
\label{eq:gamma_eta}
\gamma = 2\frac{\alpha}{\eta}+1.
\end{equation}

More importantly, one can show that as a result of $\zeta\to\infty$, the expression for hyperbolic distance $x$ between two nodes in Eq.~(\ref{eq:exact_distance}) degenerates to
\begin{equation}
x = r + r',
\end{equation}
meaning that in the $T\to\infty$ regime the angular coordinates are completely ignored.
The connection probability becomes
\begin{equation}
\label{eq:SCM_px}
p(r,r') = \frac{1}{1 + e ^ {(\eta/2) (r + r' - R)}},
\end{equation}
and the expected average degree in the network is
\begin{equation}
\bar{k} = N \int_0^R \rho(r) \int_0^R \rho(r')\; p(r,r') \; dr' dr.
\label{eq:SCM_kbar}
\end{equation}

The value of $R$ is determined using the same combination of the bisection method and numeric integration as in the previous section, except it is applied to Eq.~(\ref{eq:SCM_kbar}).

\section{Regimes with infinite $\gamma \to \infty$}\label{sec:infinite_gamma}

While in the $T\to\infty$ limit the angular coordinates are ignored, in the $\gamma \to \infty$ limit the radial coordinates are ignored. One can show this formally by observing that in this limit the radial node density approaches a delta function---all nodes are placed at the boundary at infinity of the hyperbolic plane, meaning that only angular coordinates determine distances between nodes.

\subsection{$T \in (0,\infty)$: Soft spherical random geometric graphs}

In this most general case with infinite $\gamma$, one can show that the connection probability in Eq.~(\ref{eq:default_px}) degenerates to
\begin{equation}
\label{eq:soft_RGG_px}
p(\theta, \theta') = \frac{1}{1 + \lambda\left(\frac{\Delta\theta}{\pi} \right)^\beta},
\end{equation}
where $\Delta \theta\ = \pi - | \pi - | \theta - \theta'||$ is the angular distance between the two nodes as before, while $\lambda$ is a parameter controlling the average degree $\bar{k}$ in the network, analogous to $R$ in the regimes with finite $\gamma$.
Without loss of generality we can set $\theta = 0$, so that the expression for $\bar{k}$ is
\begin{equation}
\label{eq:soft_RGG_kbar}
\bar{k} = \frac{N}{\pi} \int_0^\pi \frac{1}{1 + \lambda\left(\frac{\theta'}{\pi} \right)^\beta}\; d\theta'
=N\;{}_2F_{1}(1,T;T+1;-\lambda),
\end{equation}
where ${}_2F_1$ is the Gauss hypergeometric function, and $T=1/\beta$.
In the special case with $T=1$, the last expression simplifies to
\begin{equation}
\label{eq:soft_RGG_T1}
\frac{\bar{k}}{N}=\frac{\log(1+\lambda)}{\lambda}.
\end{equation}
If $T \neq 1$, the hypergeometric function in Eq.~(\ref{eq:soft_RGG_kbar}) cannot be evaluated using the GSL library, because the ${}_2F_1$ evaluation in the library is implemented only for the case where the fourth argument of the function ($-\lambda$ in Eq.~(\ref{eq:soft_RGG_kbar})) is between $-1$ and $1$, while for sufficiently large $N/\bar{k}$, $\lambda$ is always larger than $1$ in Eq.~(\ref{eq:soft_RGG_kbar}). To avoid this difficulty, the following transformation is used~\cite{Forrey1997}:
\begin{equation}
  \label{eq:final_softRGG}
  \begin{split}
    \frac{\bar{k}}{N} &= {}_2F_1{}(1,T;T+1;-\lambda)\\
    &= \frac{1}{\lambda+1}\frac{T}{T-1}\, {}_2F_1{}(1,1;2-T;\frac{1}{\lambda+1}) + \frac{1}{\lambda^T} \frac{\pi T}{\sin \pi T},
  \end{split}
\end{equation}
If $T > 1$ is an integer, the second term in (\ref{eq:final_softRGG}) diverges due to the $\sin$ function in the denominator, while the first term diverges because the third parameter of the ${}_2F_1{}$ function is a non-positive integer. Hence, for integer values of temperature $T>1$, their value is approximated by $T+\epsilon$, where $\epsilon$ is set to $10^{-6}$ by default. The error caused by this approximation is negligible. Equation~(\ref{eq:final_softRGG}) (or (\ref{eq:soft_RGG_T1}) if $T=1$) is then numerically solved for $\lambda$ using the bisection method, yielding the target value of $\bar{k}$ in Eq.~(\ref{eq:soft_RGG_kbar}).

\begin{table*}[ht]
  \begin{subtable}{\textwidth}
    \centering
    \begin{tabular}{|c|>{\centering}p{1.8cm}|>{\centering}p{1.8cm}|>{\centering}p{1.8cm}|>{\centering}p{1.8cm}|>{\centering\arraybackslash}p{1.8cm}|}\hline
\diaghead{zzzzzzzzzzzz}%
{$\,\gamma$}{$T$}&
$0$ & $0.5$ & $1$ & $2$ & $\infty$\\\hline
$2$& 9.66$\pm$2.19 & 9.90$\pm$2.16 & 9.98$\pm$1.08 & 9.88$\pm$1.58  & 9.84$\pm$1.47\\\hline
$3$& 10.08$\pm$0.11 & 10.01$\pm$0.10 & 10.02$\pm$0.04 & 10.00$\pm$0.08 & 9.99$\pm$0.09\\\hline
$\infty$& 10.00$\pm$0.00 &10.00$\pm$0.00 &10.00$\pm$0.00 &10.00$\pm$0.00 &10.00$\pm$0.00 \\\hline
    \end{tabular}
\caption{
Average degree.
\label{tab:kbar}
}
\vspace{1cm}
 \end{subtable}
 \begin{subtable}{\textwidth}
      \centering
    \begin{tabular}{|c|>{\centering}p{1.8cm}|>{\centering}p{1.8cm}|>{\centering}p{1.8cm}|>{\centering}p{1.8cm}|>{\centering\arraybackslash}p{1.8cm}|}\hline
\diaghead{zzzzzzzzzzzz}%
{$\,\gamma$}{$T$}&
$0$ & $0.5$ & $1$ & $2$ & $\infty$\\\hline
$2$& 0.88$\pm$0.00 & 0.75$\pm$0.00 & 0.29$\pm$0.00 & 0.36$\pm$0.01  & 0.36$\pm$0.01\\\hline
$3$& 0.79$\pm$0.00 & 0.41$\pm$0.00 & 0.42$\pm$0.00 & 0.01$\pm$0.00 & 0.01$\pm$0.00\\\hline
$\infty$& 0.75$\pm$0.00 &0.33$\pm$0.00 &0.30$\pm$0.00 &0.00$\pm$0.00 &0.00$\pm$0.00 \\\hline
\end{tabular}
\caption{
Average clustering.
\label{tab:cbar}
}
  \end{subtable}
 \caption{Observed properties in generated graphs (mean $\pm$ std\_dev for $10^3$ graph samples) with target average degree $\bar{k}=10$.}
 \label{tab:properties}
\end{table*}

\subsection{Limit $T \to 0$: Spherical random geometric graphs}\label{sec:angular_rgg}
One can see from Eq.~(\ref{eq:final_softRGG}) that the solution for $\lambda$ at small $T\ll1$ scales with $N/\bar{k}$ as $\lambda=(N/\bar{k})^\beta$, $\beta=1/T$. Therefore for $\beta\gg1$ the connection probability in Eq.~(\ref{eq:soft_RGG_px}) can be written as
\begin{equation}
p(\theta,\theta') = \frac{1}{1 + \left(\frac{N}{\bar{k}}\frac{\Delta\theta}{\pi} \right)^\beta},
\end{equation}
which in the $\beta\to\infty$ limit becomes
\begin{equation}
p(\theta,\theta') = \Theta\left(1-\frac{N \Delta \theta}{\bar{k}\pi}\right),
\end{equation}
meaning that two nodes are connected if the angular distance $\Delta\theta$ between them is smaller than $\pi\bar{k}/N$,
\begin{equation}
\label{eq:angular_RGG_threshold}
\Delta \theta < \pi\frac{\bar{k}}{N},
\end{equation}
or they are not connected otherwise.
This connectivity threshold ensures that the expected average degree in the network is $\bar{k}$.

\subsection{Limit $T\to\infty$: Erd\H{o}s-R\'enyi graphs}\label{sec:er}

In this most degenerate regime, both angular and radial coordinates are completely ignored.
This regime is formally achieved by keeping both $\alpha$ and $\zeta$ finite while letting $T \to \infty$.
One can then show that the connection probability in Eq.~(\ref{eq:default_px}) degenerates to
\begin{equation}
\label{eq:er_px}
p(x) = \frac{1}{1 + \frac{N}{\bar{k}} },
\end{equation}
which for sparse graphs with $\bar{k} \ll N$ tends to $p(x) = \bar{k} / N$, i.e., the connection probability in classical (Erd\H{o}s-R\'enyi) random graphs.

\section{Simulations}\label{sec:simulations}

Tables~\ref{tab:kbar} and~\ref{tab:cbar} show, respectively, the average degree and clustering values
in generated graphs for different regimes, $10^3$ samples in each regime. All the regimes match
the target average degree $\bar{k} = 10$, although low values of $\gamma$ lead to much
higher fluctuations. For any $\gamma>2$, average clustering decreases with temperature from a maximum
at $T=0$ to zero at $T\to\infty$.

\begin{figure}[]
  \begin{subfigure}[b]{0.5\textwidth}
    \includegraphics[width=\textwidth]{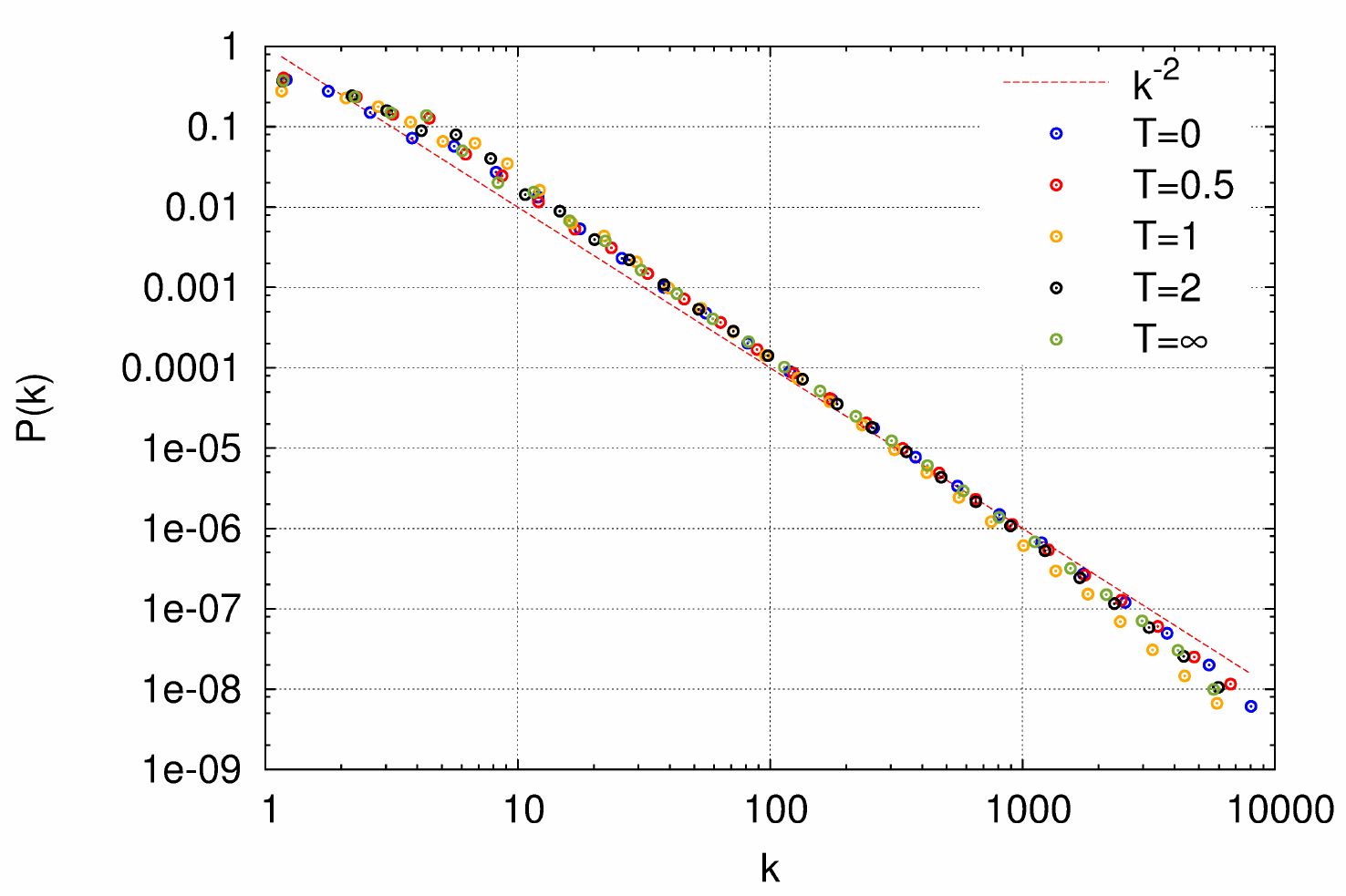}
    \caption{$\gamma = 2$}
    \label{fig:degreeDist-2}
  \end{subfigure}%

  \begin{subfigure}[b]{0.5\textwidth}
    \includegraphics[width=\textwidth]{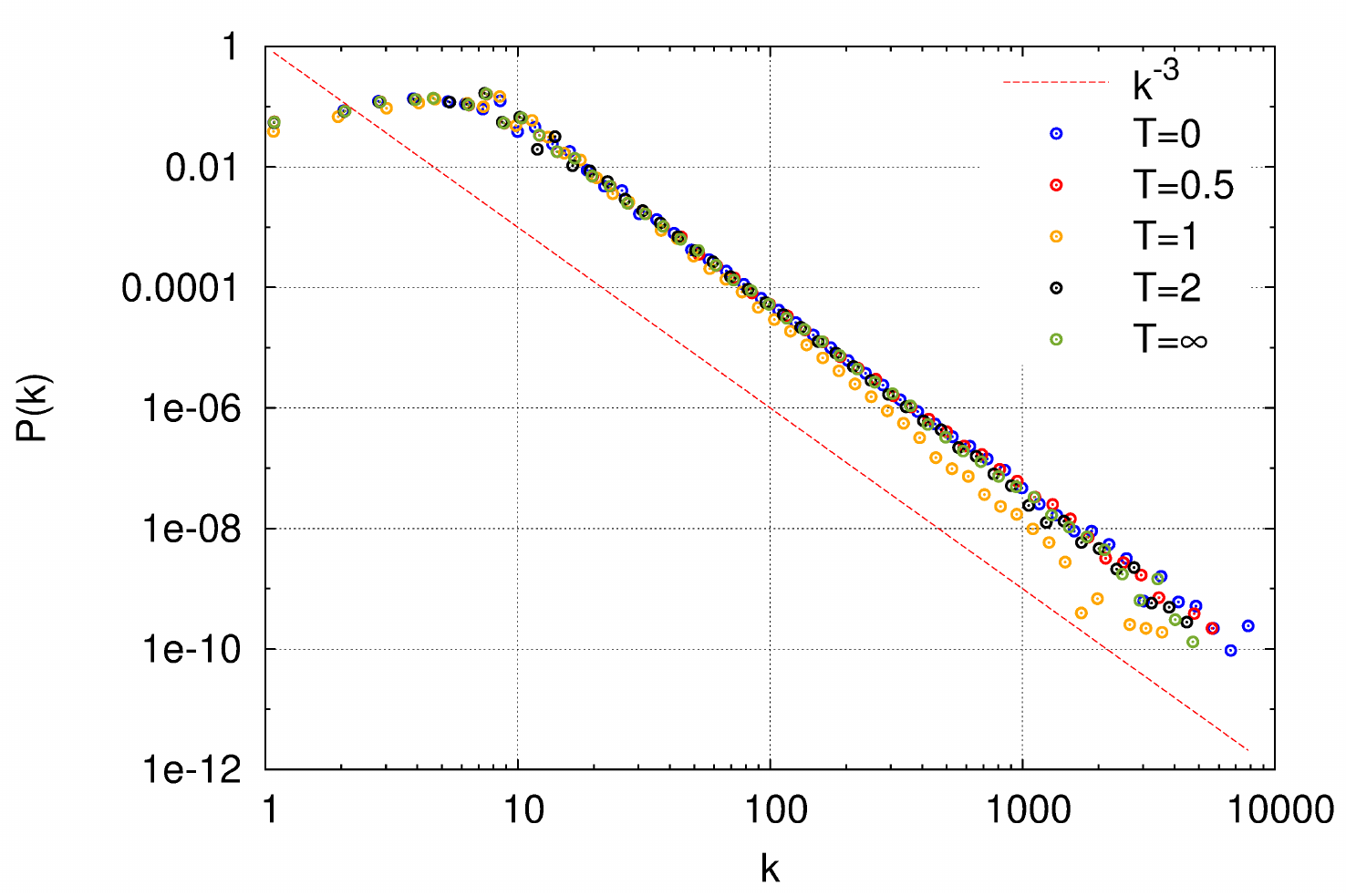}
    \caption{$\gamma = 3$}
    \label{fig:degreeDist-3}
  \end{subfigure}

  \begin{subfigure}[b]{0.5\textwidth}
    \includegraphics[width=\textwidth]{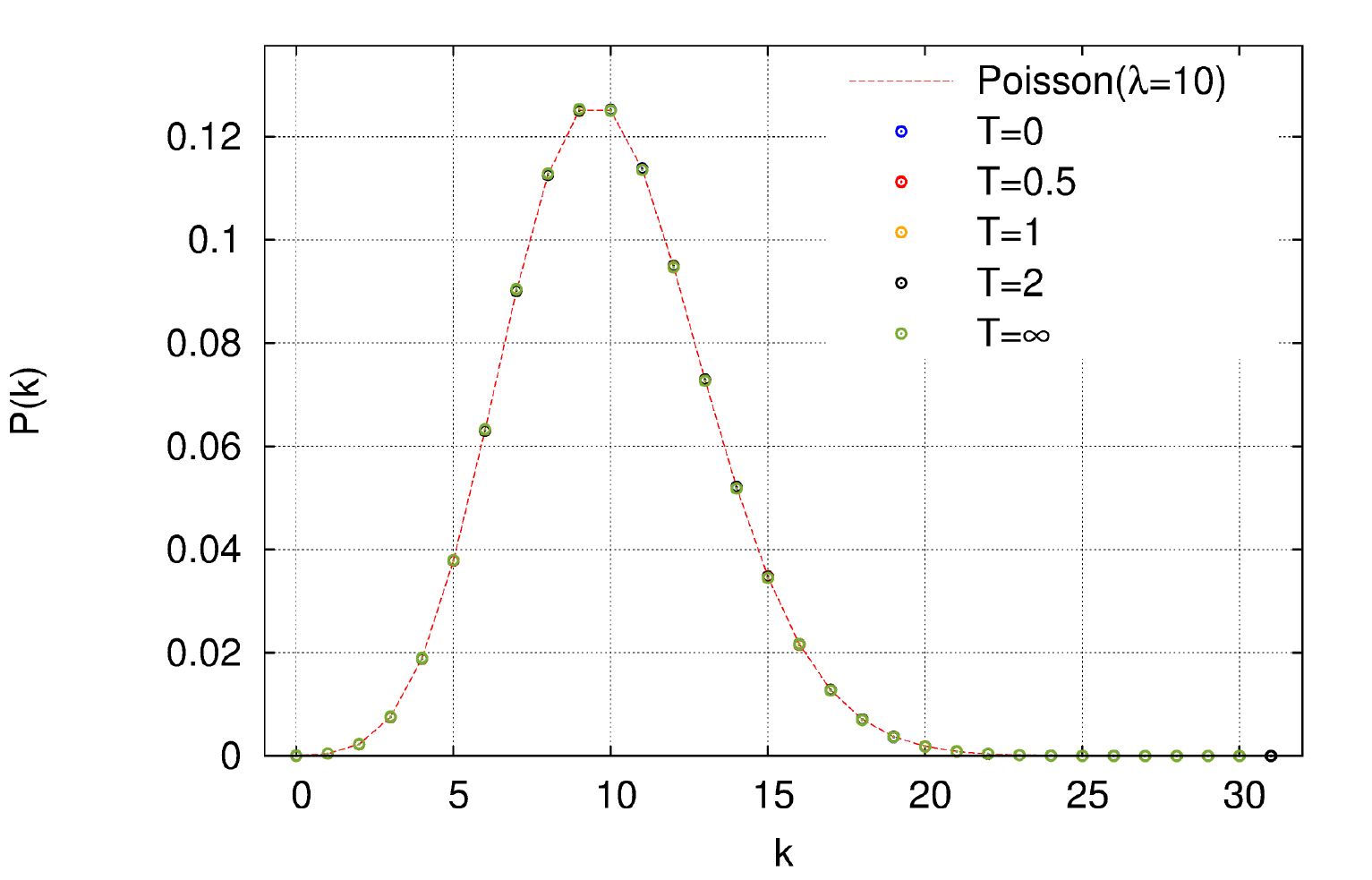}
    \caption{$\gamma = \infty$}
    \label{fig:degreeDist-inf}
  \end{subfigure}

  \caption{Observed degree distributions in generated networks of size $N=10^4$
    and target average degree $\bar{k}=10$. The results are averaged across $10^3$ generated graphs for each combinations of the parameters.}
  \label{fig:simulations-degreeDist}
\end{figure}

Figure~\ref{fig:simulations-degreeDist} shows the observed
distribution of node degrees for three different values of $\gamma$. As expected, for finite $\gamma$,
distributions follow a power-law $P(k) \sim k^{-\gamma}$. In the limit at $\gamma \to \infty$ the
observed degree distributions follow exactly the Poisson distribution with mean $\mu = \bar{k}$.

\begin{figure}[]
  \begin{subfigure}[b]{0.5\textwidth}
    \includegraphics[width=\textwidth]{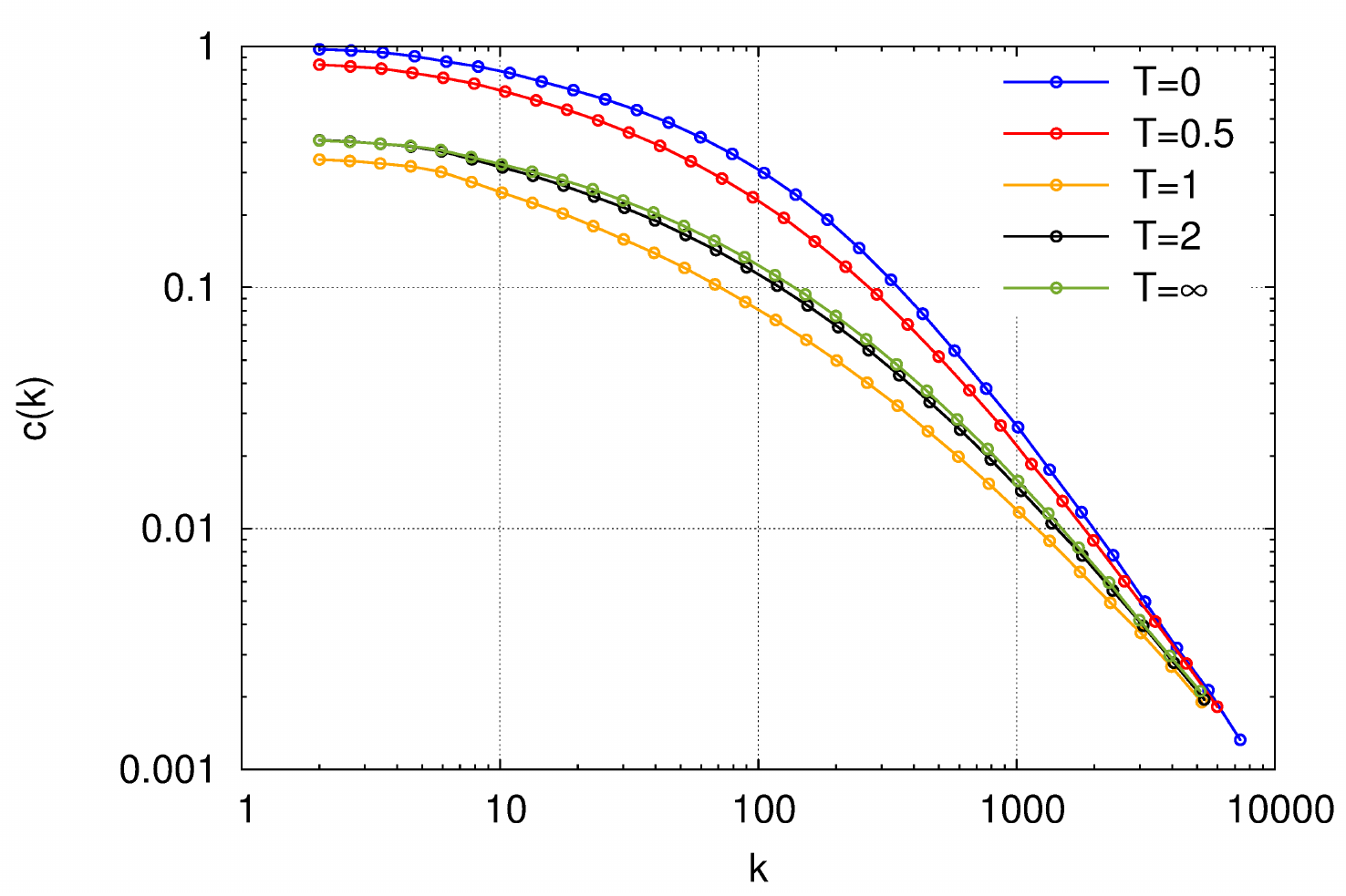}
    \caption{$\gamma = 2$}
    \label{fig:ckDist-2}
  \end{subfigure}%

  \begin{subfigure}[b]{0.5\textwidth}
    \includegraphics[width=\textwidth]{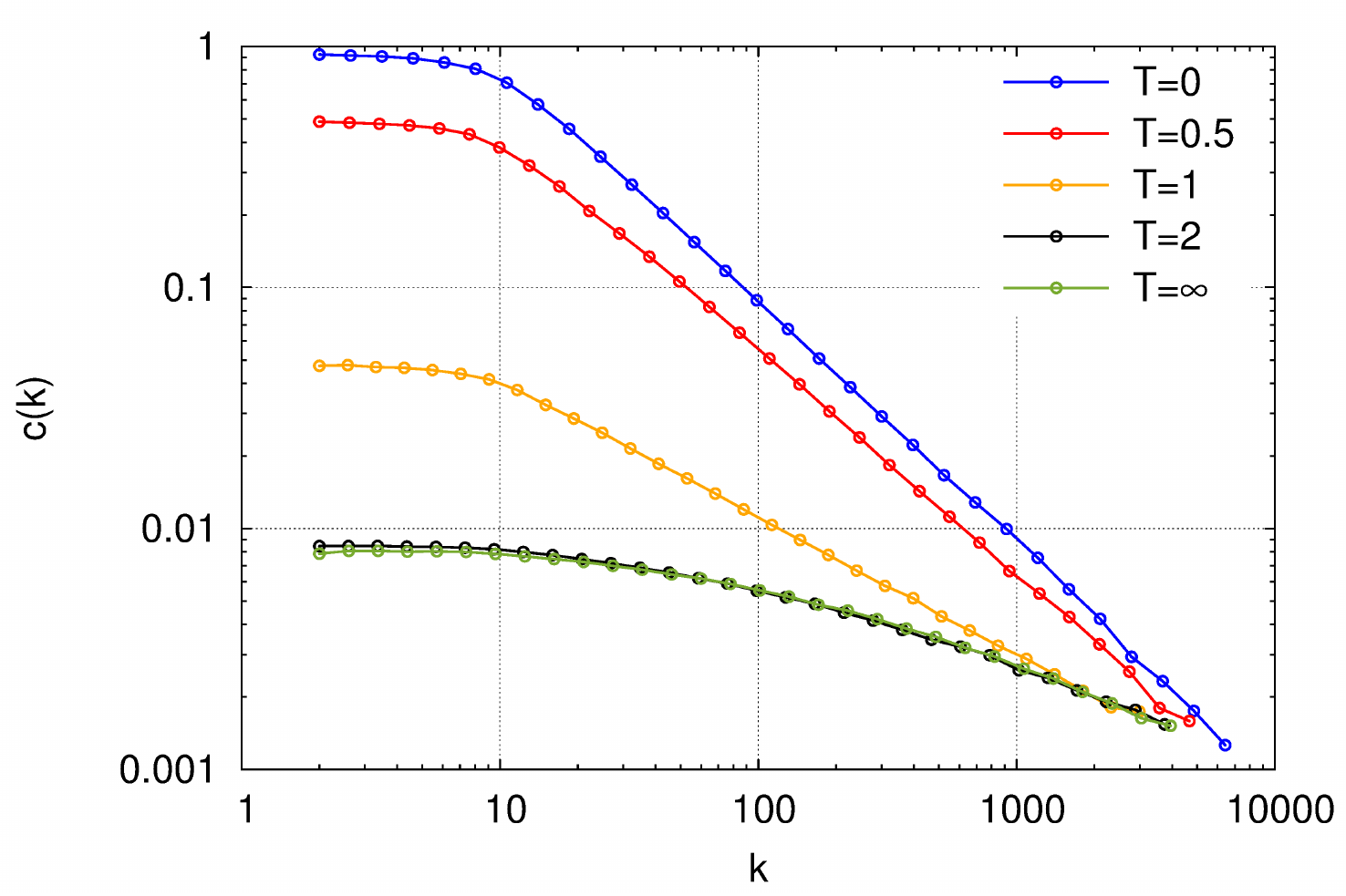}
    \caption{$\gamma = 3$}
    \label{fig:ckDist-3}
  \end{subfigure}

  \begin{subfigure}[]{0.5\textwidth}
    \includegraphics[width=\textwidth]{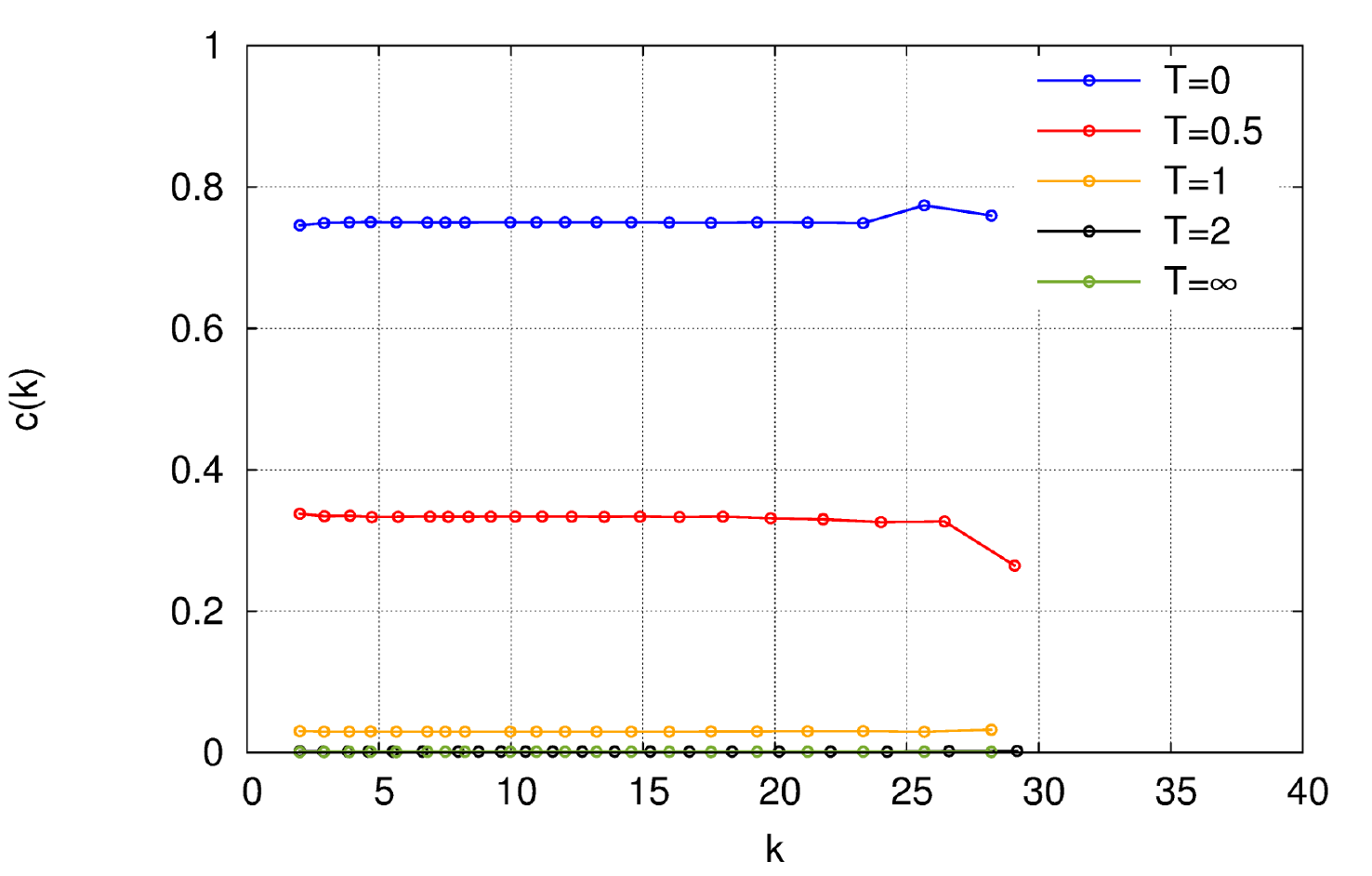}
    \caption{$\gamma = \infty$}
    \label{fig:ckDist-inf}
  \end{subfigure}
  \centering
  \caption{Observed clustering in generated networks of size $N=10^4$
    and target average degree $\bar{k}=10$. The results are averaged across $10^3$ generated graphs for each combinations of the parameters.}
  \label{fig:simulations-ckDist}
\end{figure}

Figure~\ref{fig:simulations-ckDist} shows clustering in generated graphs. For finite
$\gamma$, low-degree nodes have stronger clustering than high-degree nodes. In the case of
$\gamma \to \infty$, all nodes have similar clustering values.

\section*{Acknowledgments}

This work supported by NSF Grants No.\ CCF-1212778, CNS-1441828, and CNS-1442999.

\centering
\clearpage

\balance
\bibliographystyle{elsarticle-num-names}
\bibliography{bib}

\end{document}